\documentstyle[aps,prl,twocolumn,graphicx,floats,epsfig]{revtex}
\begin{document}
\draft
\onecolumn 

\noindent

\title{Freeze-Out Parameters in Central 158$\cdot$A~GeV 
$^{208}$Pb\/+\/$^{208}$Pb Collisions}

\author{
M.M.~Aggarwal,$^{1}$
A.~Agnihotri,$^{2}$
Z.~Ahammed,$^{3}$
A.L.S.~Angelis,$^{4}$ 
V.~Antonenko,$^{5}$ 
V.~Arefiev,$^{6}$
V.~Astakhov,$^{6}$
V.~Avdeitchikov,$^{6}$
T.C.~Awes,$^{7}$
P.V.K.S.~Baba,$^{8}$
% ...
}

\author{(WA98 Collaboration)}

\address{$^{1}$~University of Panjab, Chandigarh 160014, India}
\address{$^{2}$~University of Rajasthan, Jaipur 302004, Rajasthan,
  India}
\address{$^{3}$~Variable Energy Cyclotron Centre,  Calcutta 700 064,
  India}
\address{$^{4}$~University of Geneva, CH-1211 Geneva 4,Switzerland}
\address{$^{5}$~RRC ``Kurchatov Institute'', RU-123182 Moscow, Russia}
\address{$^{6}$~Joint Institute for Nuclear Research, RU-141980 Dubna,
  Russia}
\address{$^{7}$~Oak Ridge National Laboratory, Oak Ridge, Tennessee
  37831-6372, USA}
\address{$^{8}$~University of Jammu, Jammu 180001, India}
% ...

\date{Draft 2.0, \today}
\maketitle
\begin{abstract}
Neutral pion production in central 158$\cdot$A~GeV 
$^{208}$Pb\/+\/$^{208}$Pb  collisions has been studied in the WA98 experiment 
at the CERN SPS. The $\pi^0$ transverse mass spectrum has been
analyzed in terms of a thermal model with hydrodynamic expansion.
The high accuracy and large kinematic coverage of the measurement 
allow to limit previously noted ambiguities in the extracted
freeze-out parameters. 
The results are shown to be sensitive to the shape of the velocity 
distribution at freeze-out.

\end{abstract}
\twocolumn

Heavy ion reactions at sufficiently high energies produce dense matter 
which may provide the necessary conditions for the transition from a hadronic 
state to a deconfined phase, the Quark-Gluon Plasma. 
Since a finite thermalized system without external containment 
pressure will necessarily expand,  part of the thermal excitation 
energy will be converted into collective motion which will be 
reflected in the momentum spectra of the final hadrons.
The dynamics of the 
expansion may depend on the presence or absence of a 
plasma phase.
The strongly interacting hadrons are expected to decouple 
in the late stages of the collision. Their transverse momentum 
spectra should therefore provide information about the conditions of 
the system at freeze-out, in particular about the temperature
and collective velocity of the system, if the thermal assumption is valid.

The application of a thermal description is non-trivial. 
There is no reason to believe neither that chemical and kinetic freeze-out 
should be identical, nor that there should be  
unique thermal freeze-out temperatures for all hadrons, nor unique 
chemical freeze-out temperatures for all flavour changing reactions. 
It is likely that chemical equilibrium is not fully attained
(see e.g.~\cite{becattini96}), implying that chemical parameters 
will also influence momentum spectra through contributions from  
decays of heavier resonances.
Furthermore, it is not obvious that this problem should have  
a stationary solution since particle
emission will occur throughout the full time evolution of the
collision and so, in principle, would require a full space-time 
integration with varying parameters. 

Most attempts to extract 
freeze-out parameters from experiment assume local thermal 
equilibrium and fit parameterizations of hydrodynamical models to 
the experimental 
distributions~\cite{Schn93,Chap95,PBM95,wiedemann96,na49hydro,wa80:pi0:98}. 
Already the earliest
analyses~\cite{Schn93} noted ambiguities in fitting the hadron
transverse mass spectra due to an anti-correlation between the fitted
temperature, T,  and transverse flow velocity, $\beta_T$. 

Two-particle interferometric (HBT) measurements provide information
on the spatial and temporal extent of the emission volume, but are
also sensitive to the collective motion of the source 
(see e.g. \cite{Chap95,Pratt86,Heinz96}). Within a hydrodynamical 
parameterization of the source at freeze-out, 
the transverse two-particle correlations 
have been shown to be sensitive
only to the ratio 
$\beta_T^2/T$ \cite{Chap95}. 
Hence HBT analyses have a $\beta_T - T$ ambiguity
which is roughly orthogonal to that resulting from fits to the single
particle spectra. This fact has recently been used by the NA49
collaboration to constrain the freeze-out parameters
to lie within the region 
$\langle\beta_T\rangle=0.55\pm 0.12$ and $T=120\pm 12$ MeV 
for central Pb\/+\/Pb collisions~\cite{na49hydro}. 
Alternatively, a recent analysis
of $\pi^+,K^+,$ and $K^-$ distributions  and $\pi^+$ and
$\pi^-$ two-particle correlations measured by the NA44 collaboration 
for central  Pb\/+\/Pb collisions using a 9-parameter hydrodynamical 
model fit \cite{na44:nix:98} gave freeze-out parameters of 
$\langle \beta_T \rangle=0.443\pm 0.023$ and $T=95.8\pm 3.5$ MeV. 
These analyses
suggest that a single set of freeze-out parameters can describe the
hadron single particle distributions and two-particle correlations,
with moderate temperature and large transverse flow velocity.

On the other hand, various thermal model  analyses of particle production
ratios, especially strangeness production (see
e.g. Ref.~\cite{Sollfrank97} for a recent summary), have indicated
rather high chemical freeze-out temperatures. 
Use of integrated yields in these analyses 
allows to obtain conclusions on the temperature
which are insensitive to the amount of transverse flow.
In a recent analysis of
results at SPS energies, including Pb\/+\/Pb collisions, good
agreement is obtained if partial strangeness saturation is 
assumed with a chemical
freeze-out temperature of about 180 MeV \cite{Becattini98}. 

A successful thermal interpretation of relativistic heavy ion 
collisions must provide an accurate description of the pion spectra
since pions provide the ``thermal bath'' of the late stages
the collision.
In this letter we discuss the 
extraction of thermal freeze-out parameters from 
the neutral pion transverse mass distribution 
for central 158$\cdot$A~GeV $^{208}$Pb\/+\/$^{208}$Pb collisions.
These data provide important constraints 
due to their accuracy 
and coverage in transverse mass. 
The analysis of the $\pi^0$ spectrum, within a particular
hydrodynamical model, reveals the importance of the shape of the 
velocity distribution at freeze-out. 
The default shape, derived from a Gaussian spatial 
distribution, favors a large thermal freeze-out temperature, similar to 
temperatures extracted for chemical freeze-out, 
but in contradiction to conclusions
obtained based on analyses of limited coverage particle spectra and
HBT results \cite{na49hydro,na44:nix:98,nix:98}.

The CERN experiment WA98 
\cite{misc:wa98:proposal:91,Peitzmann:1996:qm96,wa98:pi0:98} 
consists
of large acceptance photon and hadron spectrometers together with several 
other large acceptance devices which allow to measure various global
variables on an event-by-event basis. 
The results presented here were obtained from an analysis of the
data taken with Pb beams in 1995 and 1996.
The 10\% most central reactions 
($\sigma_{central} \approx 630 \, \mathrm{mb}$) 
have been selected using the transverse 
energy $E_{T}$ measured in the MIRAC calorimeter. 

Neutral pions are reconstructed via their $\gamma\gamma$ decay branch 
using the WA98 lead-glass photon detector,
LEDA, which consisted of 10,080 individual modules with 
photomultiplier readout. The detector was located at a distance of 
21.5~m from the target and covered the pseudorapidity interval 
$2.35 < \eta < 2.95$.
The general analysis procedure, described in \cite{wa98:pi0:98},
is similar to that used in the 
WA80 experiment \cite{wa80:pi0:98}. 
The momentum 
distributions are fully corrected for geometrical acceptance and  
reconstruction efficiency. 
The systematic error on the absolute yield is $\approx 10 \%$ and
increases sharply below $p_{T}  = 0.4 \, {\mathrm{GeV}}/c$.
An additional systematic error originates 
from the uncertainty on the momentum scale of 1\%. The influence of 
this rises slowly for large $p_{T}$ and leads to an uncertainty on
the yield of 15\% 
at $p_{T} = 4 \, {\mathrm{GeV}}/c$. 

\begin{figure}[bt]
        \centerline{\includegraphics{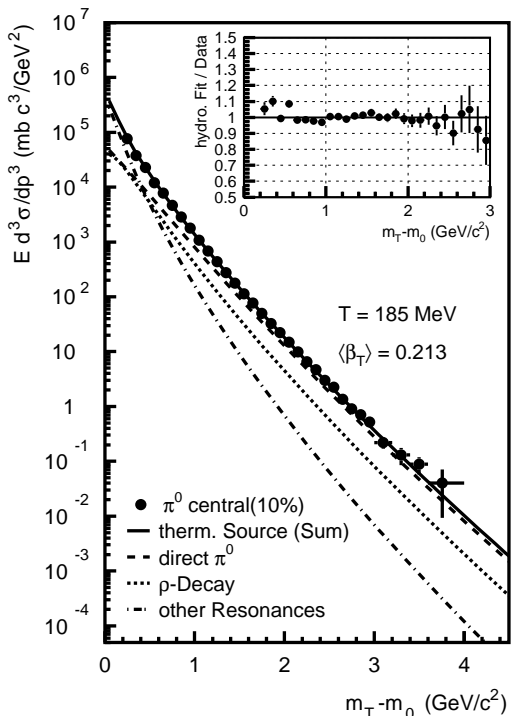}}
        \caption{Transverse mass spectra of neutral pions in central 
        collisions (10\% of min.bias cross section) of 158 $A$GeV Pb+Pb. The invariant 
        cross section of neutral pions is compared to a fit 
                using a hydrodynamical model 
        \protect\cite{wiedemann96} including transverse flow and resonance 
        decays, with the direct production and the contribution of 
        $\rho$ decays and all other resonances shown separately. 
                The ratio of the fit to the data is shown in the inset. 
                $m_0$ is the $\pi^0$ mass.
                }
        \protect\label{fig:spectra}
\end{figure}

The measured neutral pion cross section from  
central Pb+Pb reactions as a function 
of $m_{T} - m_{0}$ is shown in Fig.~\ref{fig:spectra}. 
Included is a fit with a hydrodynamical model \cite{wiedemann96} including 
transverse flow and resonance decays. This computer program
calculates the direct production and the 
contributions from the most important resonances having two- or three-body 
decays including pions ($\rho$, $\mathrm{K}^{0}_{S}$, 
$\mathrm{K}^{\star}$, $\Delta$, $\Sigma + \Lambda$, $\eta$, $\omega$, 
$\eta^{\prime}$).  The code, originally 
intended for charged pions, has been adapted to predict neutral pion
production. 
The model uses a gaussian transverse spatial density profile 
truncated at $4\sigma$. The transverse 
flow rapidity is assumed to be a linear function of the radius.
For all results presented here, a baryonic chemical 
potential of $\mu_{B} = 200 \, \mathrm{MeV}$ has been used. 
The results are 
not very sensitive, however, to the choice of $\mu_{B}$ for the $m_{T} - m_{0}$
region considered here.

This model provides an excellent description of the neutral pion 
spectra with a temperature $T = 185 \, \mathrm{MeV}$ and an average flow velocity of 
$\langle \beta_{T} \rangle = 0.213$. These values are very 
similar to the parameters obtained with similar fits to neutral pion 
spectra in central reactions of $^{32}$S+Au \cite{wa80:pi0:98}. 
The $2 \sigma$
lower limit\footnote{All limits given use the data for 
$m_{T} - m_{0} > 2 \, {\mathrm{GeV}}/c^{2}$ as upper limits only to 
allow for additional hard-scattering contributions.}
on the temperature is $T^{low} = 171 \, \mathrm{MeV}$ and 
the corresponding upper limit on the flow velocity
is $\langle \beta_{T}^{upp} \rangle = 0.253$.

The observed curvature at low $m_{T}$ is 
largely a result of resonance decay contributions. 
Performing a fit 
with only the direct contribution leads to $T = 142 \, \mathrm{MeV}$ and
$\langle \beta_{T} \rangle = 0.301$, with corresponding $2 \sigma$ 
limits of $T^{low} = 135 \, \mathrm{MeV}$ and 
$\langle \beta_{T}^{upp} \rangle = 0.318$, similar to other analyses
which have neglected decay contributions \cite{na49hydro,NA44b}. 
The larger average velocity
which results in this case is due 
to the fact that all of the observed curvature 
must now be accounted for by transverse flow. 

\begin{figure}[tbp]
        \centerline{\includegraphics[]{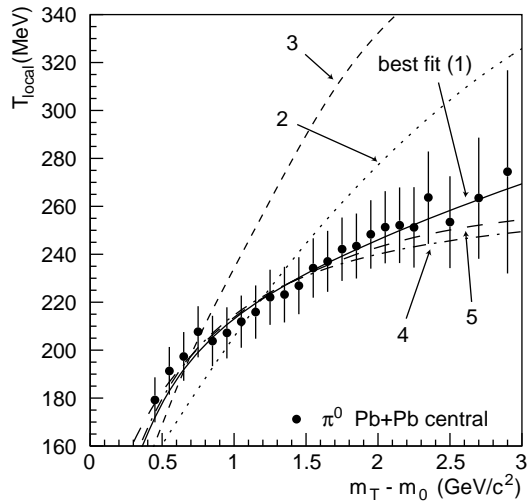}}
        \caption{The local inverse slope of the 
          transverse mass spectrum of neutral pions in central 
        collisions of 158 $A$GeV Pb+Pb. The measured results (solid points)
        are compared to the hydrodynamical model best fit result 
        (solid line; $T = 185 \, \mathrm{MeV}$ and 
        $\langle \beta_{T} \rangle = 0.213$, also shown in Fig.~1) 
        and to the other results given in table \ref{tbl:para}.
        }
        \protect\label{fig:slope}
\end{figure}

The high statistical accuracy and large transverse mass coverage of the
present $\pi^0$ measurement reveals 
the concave curvature of the $\pi^0$ spectrum  
over a large $m_T$ range, which constrains the parameters significantly. 
This is further demonstrated by studying the
local slope at each $m_T$. 
The local (inverse) slope is given by

\begin{equation}
T^{-1}_{local} = -\left( E\frac{d^3\sigma}{dp^3}\right)^{-1}\frac{d}{dm_T}
\left( E\frac{d^3\sigma}{dp^3} \right).
\end{equation}

The local slope results are plotted in  Fig.~\ref{fig:slope}. Each individual
value of $T_{local}$ has been extracted from 3 adjacent data points
of Fig.~\ref{fig:spectra}. The data are compared to the hydrodynamical
model best fit results of Fig.~\ref{fig:spectra}, 
as well as fits
in which the 
transverse flow velocities have been fixed to larger 
values comparable to those obtained
by Refs.~\cite{na44:nix:98} 
and NA49~\cite{na49hydro} (sets 2 and 3). The 
corresponding fit parameters are given in Table~\ref{tbl:para}.
The comparison demonstrates
that while the large transverse flow velocity fits can provide a reasonable
description of the data up to transverse masses of about 1 GeV, they 
significantly overpredict the local slopes at large transverse mass. 
While application of the hydrodynamical model at large transverse mass
is questionable, the model cannot overpredict the measured yield. 
The observed overprediction therefore rules out the assumption of
large transverse flow velocities, or points to a deficiency in the 
model assumptions used in these fits. 

\begin{figure}[tbp]
        \centerline{\includegraphics[]{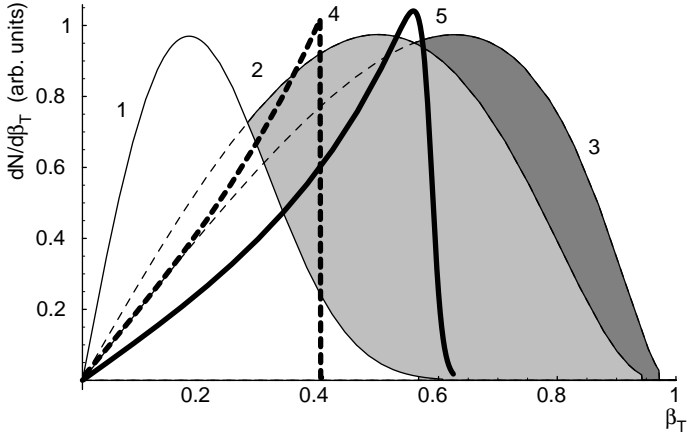}}
        \caption{Unnormalized multiplicity distributions as a 
        function of the transverse source velocity 
        for the parameter sets given 
        in table \ref{tbl:para}.
        }
        \protect\label{fig:profile}
\end{figure}

The curvature in the $\pi^0$ spectrum at large transverse mass is a
result of the distribution of transverse velocities. Although the
spectrum is not directly sensitive to the spatial distribution of
particle emission, within this model it is dependent indirectly on 
the spatial distribution due to the assumption that the 
transverse rapidity increases linearly with radius.
The large curvature at large transverse mass is due to 
high velocity contributions which result from the tail of
the assumed gaussian density profile~\cite{Heinz98}.
Figure \ref{fig:profile} shows the transverse source velocity 
distributions $dN/d\beta_T$ for the different parameter sets. 
\footnote{More precisely these are source emission functions 
integrated over all variables except the transverse velocity and the 
rapidity, i.e. they are
weighted with the produced particle multiplicity.}
The curves labelled 1-3 
correspond to the calculations in figure \ref{fig:slope} using a 
gaussian spatial profile. In addition, velocity profiles are shown
for a uniform density profile 
(set 4) and for a Woods-Saxon distribution:
\begin{equation}
    \rho(r) = \frac{1}{1 + \exp\left[ (r - r_{0}) / \Delta \right]}
    \label{eq:ws}
\end{equation}
with $\Delta/r_{0} = 0.02$ (set 5). These are 
included in figures \ref{fig:slope} and \ref{fig:profile}.
It is seen that the uniform density assumption truncates the high
velocity tail resulting in less curvature in the pion spectrum, while the 
Woods-Saxon has a more diffuse edge at high $\beta_T$.

While the gaussian and uniform density assumptions have very different
velocity profiles, it is interesting that both can provide acceptable
fits to the pion spectrum with best fit results with
similar $\langle \beta_{T}
\rangle$ and $T$ parameters, which give similar effective
temperatures, and which have similar velocity widths,  $\beta_{RMS}$, as shown in
Table~\ref{tbl:para}.
Compared to the gaussian profile result, 
the best fit result using the uniform profile 
gives a lower temperature of 178 MeV and would lead to weaker limits 
of $\langle \beta_{T}^{upp} \rangle = 0.42$ 
and $T^{low} = 134 \, \mathrm{MeV}$.
Limits cannot be set using the Woods-Saxon profile due to increased
fit ambiguity. If the data for 
\mbox{$m_{T} - m_{0} > 2 \,{\mathrm{GeV}}/c^{2}$} 
is used only as upper
limits, as explained above, a best fit result with 
$T = 129 \, \mathrm{MeV}$ and
$\langle \beta_{T} \rangle = 0.42$ is obtained.
 
%
The data presented here can be well described 
with high thermal freeze-out temperatures, 
similar to temperatures which have been extracted for chemical freeze-out
\cite{Becattini98}, and small transverse flow velocities. 
\footnote{Note again that chemical and thermal freeze-out
are not necessarily expected to be the same.}
On the other hand, if the larger velocities
obtained in other
analyses which have considered limited particle spectra together with
HBT results~\cite{na49hydro,na44:nix:98,nix:98} persist,
then the present analysis suggests much lower thermal
freeze-out temperatures.
For example, none of the different velocity
profile assumptions used in this analysis allowed 
to reproduce the results of 
ref.~\cite{na49hydro} -- all profiles studied require a temperature of 
90~MeV or less, if $\langle \beta_{T} \rangle = 0.55$ is assumed.
The present data obviously provide important information 
on the shape of the freeze-out velocity distribution.
A more extensive systematic study 
would require further guidance 
from full hydrodynamical calculations, which is beyond the scope of this
paper. 
Recent hydrodynamical model calculations~\cite{Dumitru98,Huovinen98}
have found reasonable agreement with transverse mass spectra within a 
broad range of assumptions. However, in these studies 
it was not attempted to limit the model 
parameters or assumptions by a rigorous comparison to the data. 

\begin{table}[ht]
    \centering
    \caption{Parameters for different hydrodynamical model fits to the 
    neutral pion spectrum shown in figures \ref{fig:slope} and 
    \ref{fig:profile}. The temperature $T$,  average and RMS transverse flow 
    velocity $\langle \beta_{T} \rangle$ and $\beta_{RMS}$
    are given together with the effective temperature 
    $T_{eff} = T/ \sqrt{(1-\langle \beta_T \rangle)/
    (1+\langle \beta_T \rangle)}$. }
    \begin{tabular}{|c|c|rl|rl|r|r|r|}
        Set & spatial & \multicolumn{2}{c|}{$T$} 
        & \multicolumn{2}{c|}{$\langle \beta_{T} \rangle$} & 
        $\beta_{RMS}$ & $T_{eff}$ & $\chi^{2}/dof$ \\
         &  profile & \multicolumn{2}{c|}{(MeV)} 
        & \multicolumn{2}{c|}{} & 
         &(MeV) & \\
        \hline
        1 & Gauss & $185$ & $\pm 4$ & $0.213$ & $\pm 0.020$ & 0.107 & 
        230 & 25.9/18 \\
        \hline
        2 & Gauss & $75$ & $\pm 1$ & 0.469 & & 0.199 & 125 & 386/19 \\
        \hline
        3 & Gauss & $49$ & $\pm 1$ & 0.527 & & 0.213 & 88 & 578/19 \\
        \hline
        4 & Uniform & $178$ & $\pm 13$ & $0.274$ & $\pm 0.046$ 
              & 0.093 & 235 & 33.3/18 \\
        \hline
        5 & WS & $146$ & $^{+21}_{-16}$ & $0.365$ & $^{+0.056}_{-0.069}$ 
	& 0.137 & 214 & 26.7/18 \\
    \end{tabular}
    \label{tbl:para}
\end{table}

In summary, we have argued that
hydrodynamical models which attempt to extract the thermal freeze-out 
parameters of relativistic heavy ion collisions must 
provide an accurate description of the pion spectra,
since pions most directly reflect the thermal evironment
in the late stage of the collision.
In particular, models, or parameter sets, which overpredict
the observed pion yields, even at large transverse mass, 
can immediately be ruled out. 
We have demonstrated that the high accuracy neutral pion 
spectra with large transverse mass coverage can constrain the 
thermal freeze-out parameters and model assumptions.
Within the context of the hydrodynamical
model of Ref.~\cite{wiedemann96}, the default velocity
profile favors large thermal freeze-out temperatures  
similar to the chemical 
freeze-out temperature determined for the same system~\cite{Becattini98}.
Only special choices of the 
velocity profile allow large average freeze-out velocities similar 
to those extracted from other recent analyses which consider also 
HBT results~\cite{na49hydro,na44:nix:98,nix:98}. On the other hand,
the corresponding freeze-out temperatures are then $\approx 90$ MeV,
significantly lower than other estimates. The present results 
indicate that the determination of the freeze-out parameters remains 
an open question. It will be important to determine whether full
hydrodynamical models can reproduce the high precision pion data
and thereby constrain the assumed freeze-out hypersurface.

We wish to thank Urs Wiedemann for assistance with the model
calculations and valuable discussions.
This work was supported jointly by 
the German BMBF and DFG, 
the U.S. DOE,
the Swedish NFR and FRN, 
the Dutch Stichting FOM, 
the Stiftung f{\"u}r Deutsch-Polnische Zusammenarbeit,
the Grant Agency of the Czech Republic under contract No. 202/95/0217,
the Department of Atomic Energy,
the Department of Science and Technology,
the Council of Scientific and Industrial Research and 
the University Grants 
Commission of the Government of India, 
the Indo-FRG Exchange Program,
the PPE division of CERN, 
the Swiss National Fund, 
the INTAS under Contract INTAS-97-0158, 
ORISE, 
Grant-in-Aid for Scientific Research
(Specially Promoted Research \& International Scientific Research)
of the Ministry of Education, Science and Culture, 
the University of Tsukuba Special Research Projects, and
the JSPS Research Fellowships for Young Scientists.
ORNL is managed by Lockheed Martin Energy Research Corporation under
contract DE-AC05-96OR22464 with the U.S. Department of Energy.
The MIT group has been supported by the US Dept. of Energy under the
cooperative agreement DE-FC02-94ER40818.

%
%


\begin{references}

\bibitem{becattini96}
F.~Becattini, Z. Phys. C {\bf 69} (1996) 485--492.

\bibitem{Schn93}
E.~Schnedermann, J.~Sollfrank, and U.~Heinz, Phys. Rev. C {\bf48} (1993) 2462.

\bibitem{Chap95} 
S. Chapman, J. R. Nix and U. Heinz, Phys. Rev. C {\bf52} (1995) 2694.

\bibitem{PBM95}
P~.Braun-Munzinger et~al., Phys. Lett. B {\bf 344} (1995) 43.

\bibitem{wiedemann96}
U.A.~Wiedemann and U.~Heinz, Phys. Rev. C {\bf 56} (1997) 3265.

\bibitem{na49hydro}
NA49 Collaboration, 
H.~Appelsh\"{a}user et~al., Eur. Phys. J. C {\bf 2} (1998) 661--670.

\bibitem{wa80:pi0:98} WA80 Collaboration, 
R.~Albrecht et al., Eur. Phys. J. C {\bf 5} (1998) 255.

\bibitem{Pratt86}
S.~Pratt, Phys. Rev. D {\bf 33} (1986) 1314.

\bibitem{Heinz96} 
U. Heinz et al., Phys. Lett. B {\bf382} (1996) 181.

\bibitem{na44:nix:98} J.R.~Nix et~al.,nucl-th/9801045.

\bibitem{Sollfrank97}
J.~Sollfrank, J. Phys. G {\bf 23} (1997) 1903.

\bibitem{Becattini98}
F.~Becattini, M.~Ga{\'z}dzicki, and J.~Sollfrank, Eur. Phys. J. C {\bf
  5} (1998) 143.

\bibitem{nix:98} J.R.~Nix, Phys. Rev. C {\bf58} (1998) 2303.

\bibitem{misc:wa98:proposal:91}
WA98 Collaboration,
\newblock {\em Proposal for a large acceptance hadron and photon spectrometer},
  1991,
\newblock Preprint CERN/SPSLC 91-17, SPSLC/P260.

\bibitem{Peitzmann:1996:qm96}
WA98 Collaboration, M.~Aggarwal et~al., Nucl. Phys. A {\bf 610} (1996) 200c.

\bibitem{wa98:pi0:98} WA98 Collaboration, M.M.~Aggarwal et al.,
Phys. Rev. Lett. {\bf 81} (1998) 4087.

\bibitem{NA44b}  NA44 Collaboration, I.G.~Bearden et al., Phys. Rev. Lett. 
{\bf 78} (1997) 2080.

\bibitem{Heinz98} U.~Heinz and U.A.~Wiedemann, private communication.

\bibitem{Dumitru98} A.~Dimitru and D.H.~Rischke, Phys. Rev. {\bf C59} 
(1999) 354-363.

\bibitem{Huovinen98} P.~Huovinen, P.V.~Ruuskanen, and J.~Sollfrank, 
Nucl. Phys. {\bf A 650} (1999) 227-244.

\end{references}
\end{document}